\documentclass[floats,nobibnotes,nofootinbib,showpacs,11pt,onecolumn,nobalancelastpage,tightenlines,rmp]{revtex4}
\usepackage{amssymb}
\usepackage{epsfig}

\begin{document}

\title{Functional Optimization in Complex Excitable Networks}
\author{Samuel Johnson\thanks{samuel@onsager.ugr.es}, J. Marro, and Joaqu\'{\i}n J. Torres}
\affiliation{Departamento de Electromagnetismo y F\'{\i}sica de la Materia\textit{,} and
Institute \textquotedblleft Carlos I\textquotedblright\ for Theoretical and
Computational Physics,\\
Facultad de Ciencias, Universidad de Granada, 18071--Granada, Spain.}

\begin{abstract}
We study the effect of varying wiring in excitable random networks in which
connection weights change with activity to mold local resistance or
facilitation due to fatigue. Dynamic attractors, corresponding to patterns
of activity, are then easily destabilized according to three main modes,
including one in which the activity shows chaotic hopping among the
patterns. We describe phase transitions to this regime, and show a
monotonous dependence of critical parameters on the heterogeneity of the
wiring distribution. Such correlation between topology and functionality
implies, in particular, that tasks which require unstable behavior ---such
as pattern recognition, family discrimination and categorization--- can be
most efficiently performed on highly heterogeneous networks. It also follows
a possible explanation for the abundance in nature of scale--free network
topologies.
\end{abstract}

\maketitle
Excitable systems allow for the regeneration of waves propagating through
them, and may thus respond vigorously to weak stimulus. The brain and other
parts of the nervous system are well--studied paradigms, and forest fires
with constant ignition of trees and autocatalytic reactions in surfaces, for
instance, also share some of the basics \cite{bak,Meron,Lindner,Diaz,Izhi}.
The fact that signals are not gradually damped by friction in these cases is
a consequence of cooperativeness between many elements in a nonequilibrium
setting. In fact, the systems of interest may be viewed as large networks
whose nodes are \textquotedblleft excitable\textquotedblright . This, which
admits various realizations, typically means that each element has a
threshold and a refractory time between consecutive responses.

Brain tasks may ideally be reproduced in mathematical neural networks. These
consist of \textit{neurons} ---often simplified as binary variables---
connected by edges representing \textit{synapses }\cite{Amari,Hopfield,Amit}%
. Assuming edges weighted according to a prescription (e.g. \cite{Hebb})
which in a sense saves information from a set of given patterns of activity,
these patterns become attractors of the phase--space dynamics. Therefore,
the system shows retrieval of the stored patterns, known as \textit{%
associative memory}. Actual neural systems do much more than just recalling
a memory and staying there, however. That is, one should expect dynamic
instabilities or other destabilizing mechanism. This expectation is
reinforced by recent experiments suggesting that synapses undergo rapid
changes with time which may both determine brain tasks \cite%
{Abbot_synaptic,Tsodyks,Pantic,Malenka} and induce irregular and perhaps
chaotic activity \cite{Barrie,Korn}.

The observed rapid changes (which have been described \cite%
{Tsodyks,Pantic,Malenka} as causing \textquotedblleft synaptic
depression\textquotedblright\ and/or \textquotedblleft
facilitation\textquotedblright\ on the time scale of milliseconds ---i.e.,
much faster than the plasticity process in which synapses \textit{store}
patterns) may correspond to the characteristic behavior of single excitable
elements. Furthermore, a fully--connected network which describes
cooperation between such excitable elements exhibits both attractors and
chaotic instabilities \cite{Marro_complex}. Here, we extend and generalize
this study to conclude on the influence of the excitable network topology on
dynamic behavior. We show, in particular, an interesting correlation between
certain wiring topology and optimal functionality.

Consider $N$ binary nodes, $s_{i}=\pm 1,$ and the topology matrix, $\epsilon
_{ij}=1,0,$ which indicates the existence or not of an edge between nodes $%
i,j=1,2,...,N.$ Let a set of \textit{M} patterns, $\xi _{i}^{\nu }=\pm 1,$ $%
\nu =1,...M$ (which we generate here at random), and assume that they are
\textquotedblleft stored\textquotedblright\ by giving each edge a base
weight $\overline{\omega _{ij}}=N^{-1}\sum_{\nu }\xi _{i}^{\nu }\xi
_{j}^{\nu }$. Actual weights are dynamic, however, e.g., $\omega _{ij}=%
\overline{\omega _{ij}}x_{j}$ where $x_{j}$ is a stochastic variable.
Assuming the limit in which this varies in a time scale infinitely smaller
than the one for node dynamics, we may consider a stationary distribution
such as $P(x_{j}|S)=q\delta (x_{j}-\Xi _{j})+(1-q)\delta (x_{j}-1),$ $%
S=\left\{ s_{j}\right\} ,$ for instance. This amounts to assume that, at
each time step, every connection has a probability \textit{q} of altering
its weight by a factor $\Xi _{j}$ which is a function (to be determined) of
the local \textit{field} at $j,$ namely, the net current from other nodes.
This choice differs essentially from the one in Ref.\cite{Marro_complex},
where $q$ depends on the global degree of order and $\Xi _{j}$ is a constant
independent of $j.$%

Assume independence of the noise at different edges, and that the transition
rate for the stochastic changes is%
\[
\frac{\bar{c}\left( S\rightarrow S^{i}\right) }{\bar{c}\left(
S^{i}\rightarrow S\right) }=\prod_{j/\epsilon _{ij}=1}\frac{\int
dx_{j}P(x_{j}|S)\Psi (u_{ij})}{\int dx_{j}P(x_{j}|S^{i})\Psi (-u_{ij})}, 
\]%
where $u_{ij}\equiv s_{i}s_{j}x_{j}\overline{\omega _{ij}}T^{-1},$ $\Psi
(u)=\exp \left( -%
{\frac12}%
u\right) $ to have proper contour conditions, \textit{T} is a
\textquotedblleft temperature\textquotedblright\ parameter, and $S^{i}$
stands for \textit{S} after the change $s_{i}\rightarrow -s_{i}.$ (For a
description of this formalism and its interpretation, see \cite%
{Marro_nonequilibrium}.) We define the \textit{effective local fields} $%
h_{i}^{\text{eff}}=h_{i}^{\text{eff}}(S,T,q)$ via $\prod_{j}\varphi
_{ij}^{-}/\varphi _{ij}^{+}=\exp \left( -h_{i}^{\text{eff}}s_{i}/T\right) ,$
where $\varphi _{ij}^{\pm }\equiv q\exp \left( \pm \Xi _{j}v_{ij}\right)
+(1-q)\exp \left( \pm v_{ij}\right) $, with $v_{ij}=\frac{1}{2}\epsilon
_{ij}u_{ij}.$ Effective weights $\omega _{ij}^{\text{eff}}$ then follow from 
$h_{i}^{\text{eff}}=\sum_{j}\omega _{ij}^{\text{eff}}s_{j}\epsilon _{ij}$.
To obtain an analytical expression, we linearize around $\overline{\omega
_{ij}}=0$ (a good approximation when $M\ll N$), which yields 
\[
\omega _{ij}^{\text{eff}}=\left[ 1+q\left( \Xi _{j}-1\right) \right] 
\overline{\omega _{ij}}. 
\]%
In order to fix $\Xi _{j}$ here, we first introduce the overlap vector $%
\overrightarrow{m}=(m^{1},...m^{M}),$ with $m^{\nu }\equiv N^{-1}\sum_{i}\xi
_{i}^{\nu }s_{i},$ which measures the correlation between the current
configuration and each of the stored patterns, and the \textit{local} one $%
\overrightarrow{m_{j}}$ of components $m_{j}^{\nu }\equiv \langle k\rangle
^{-1}\sum_{l}\xi _{l}^{\nu }s_{l}\epsilon _{jl}$, where $\langle k\rangle $
is the mean node connectivity, i.e., the average of $k_{i}=\sum_{j}\epsilon
_{ij}.$ We then assume, for any $q\neq 0,$ that the factor is $\Xi
_{j}=1+\zeta (h_{j}^{\nu })(\Phi -1)/q,$ with 
\[
\zeta (h_{j}^{\nu })=\chi ^{\alpha }/\left( 1+M/N\right) \sum_{\nu
}|h_{j}^{\nu }|^{\alpha } 
\]%
where $\chi \equiv N/\langle k\rangle $ and $\alpha >0$ is a parameter. This
comes from the fact that the field at node $j$ may be written as a sum of
components from each pattern, namely, $h_{j}=\sum_{\nu }^{M}h_{j}^{\nu }$,
where 
\[
h_{j}^{\nu }=\xi _{j}^{\nu }N^{-1}\sum_{i}\epsilon _{ij}\xi _{i}^{\nu
}s_{i}=\chi ^{-1}\xi _{j}^{\nu }m_{j}^{\nu }. 
\]%
Our choice for $\Xi _{j},$ which amounts to assume that the
\textquotedblleft fatigue\textquotedblright\ at a given edge increases with
the field at the preceding node $j$ (and allows to recover the
fully--connected limit in \cite{Marro_complex} if $\alpha =2$), finally
leads to%
\[
\omega _{ij}^{\text{eff}}=\left[ 1+(\Phi -1)\zeta _{j}(\overrightarrow{m_{j}}%
)\right] \overline{\omega _{ij}}. 
\]%
Varying $\Phi $ one sets the nature of the weights. That is, $0<\Phi <1$
corresponds to resistance (\textit{depression}) due to heavy local work,
while the edge facilitates, i.e., tends to increase the effect of the signal
under the same situation for $\Phi >1.$ (The action of the edge is reversed
for negative $\Phi .)$ We performed Monte Carlo simulations using standard
parallel updating with the effective rates $\bar{c}\left( S\rightarrow
S^{i}\right) $ computed using the latter effective weights.

It is possible to solve the single pattern case ($M=1$) under a mean-field
assumption, which is a good approximation for large enough connectivity.
That is, we may substitute the matrix $\epsilon _{ij}$ by its mean value
over network realizations to obtain analytical results that are independent
of the underlying disorder. Imagine that each node hosts $k_{i}$ \textit{%
half--edges} according to a distribution $p(k),$ the total number of
half--edges in the network being $\langle k\rangle N$. Choose a node $i$ at
random and randomly join one of its half--edges to an available free
half--edge. The probability that this half--edge ends at node $j$ is $%
k_{j}/\left( \left\langle k\right\rangle N\right) .$ Once all the nodes have
been linked up, the expected value (as a quenched average over network
realizations) for the number of edges joining nodes \textit{i} and \textit{j}
is $\overline{\epsilon _{ij}}=k_{i}k_{j}/\left( \left\langle k\right\rangle
N\right) $ \cite{footnote1}. Using the notation $\eta _{i}\equiv \xi
_{i}s_{i}$, we have $m_{j}=\chi \langle \eta _{i}\epsilon _{ij}\rangle _{i}=%
\frac{\chi }{N}\sum_{i}\eta _{i}\epsilon _{ij}.$ Because node activity is
not statistically independent of connectivity \cite{Torres_influence}, we
must define a new set of overlap parameters, analogous to $m$ and $m_{j}.$
That is, $\mu _{n}\equiv \langle k_{i}^{n}\eta _{i}\rangle _{i}/\langle
k^{n}\rangle $ and the local versions $\mu _{n}^{j}\equiv \chi \langle
k_{i}^{n}\eta _{i}\epsilon _{ij}\rangle _{i}/\langle k^{n}\rangle .$ After
using $\epsilon _{ij}=\overline{\epsilon _{ij}},$ one obtains the relation $%
\mu _{n}^{i}=\langle k^{n+1}\rangle k_{i}\mu _{n+1}/(\langle k^{n}\rangle
\langle k\rangle ^{2}).$ Inserting this expression into the definition of $%
\mu _{n}$, and substituting $\langle s_{i}\rangle =\tanh
[T^{-1}h_{i}^{eff}(S)]$ (for very large N), standard mean-field analysis
yields%
\[
\mu _{n}(t+1)=\frac{1}{\left\langle k^{n}\right\rangle }\left\langle
k^{n}\tanh M_{T,\Phi }(k,t)\right\rangle _{k}, 
\]%
where the last quantity is defined as%
\[
M_{T,\Phi }=\frac{k}{TN}\left[ \mu _{1}(t)+(\Phi -1)\frac{\langle k^{\alpha
+1}\rangle }{\langle k\rangle ^{\alpha +1}}\left\vert \mu _{1}(t)\right\vert
^{\alpha }\mu _{\alpha +1}(t)\right] . 
\]%
This is a two-dimensional map which is valid for any random topology of
distribution $p(k)$. Note that the macroscopic magnitude of interest is $\mu
_{0}=m\equiv |\overrightarrow{m}|.$

A main consequence of this is the existence of a critical temperature, $%
T_{c},$ under very general conditions, e.g., for many different network
connectivities. More specifically, as $T$ is decreased, the overlap $m$
describes a second--order phase transition from a disordered or, say,
\textquotedblleft paramagnetic\textquotedblright\ phase to an ordered
(\textquotedblleft ferromagnetic\textquotedblright ) phase which exhibits
associative memory. The mean--field temperature signaling this transition is%
\[
T_{c}=\langle k^{2}\rangle \left( \left\langle k\right\rangle N\right)
^{-1}. 
\]%
On the other hand, the map reduces for $T=0$ to $\mu _{n}\left( t+1\right) =%
\mbox{sign}\left\{ \mu _{n}\left( t\right) \left[ 1+(\Phi -1)\langle
k\rangle ^{\alpha +1}/\langle k^{\alpha +1}\rangle \right] \right\} .$ This
implies the existence at $\Phi =\Phi _{0},$ where%
\[
\Phi _{0}=1-\langle k\rangle ^{\alpha +1}/\langle k^{\alpha +1}\rangle , 
\]%
of a transition as $\Phi $ is decreased from the ferromagnetic phase to a
new phase in which periodic hopping between the attractor and its negative
occurs. This is confirmed by the Monte Carlo simulations for $M>1,$ namely,
the hopping is also among different attractors for finite $T.$ The
simulations also indicate that this transition washes out at low enough
finite temperature. Instead, actual evolutions show that, for a certain
range of $\Phi $ values, the system activity then exhibits chaotic behavior.

The transition from ferromagnetic to chaotic states is a main concern
hereafter. Our interest in this regime follows from several recent
observations concerning the relevance of chaotic activity in a network. In
particular, it has been shown that chaos might be responsible for certain
states of attention during brain activity \cite{Torres_int,TorresNEW}, and
that some network properties such as the computational capacity \cite{Bert}
and the dynamic range of sensitivity to stimuli \cite{Assis} may become
optimal at the \textquotedblleft edge of chaos\textquotedblright\ in a
variety of settings.

We next notice that the critical values $T_{c}$ and $\Phi _{0}$ only depend
on the moments of the generic distribution $p(k),$ and that the ratio $%
\langle k^{a}\rangle /\langle k\rangle ^{a},$ $a>1$, is a convenient way of
characterizing heterogeneity. We studied in detail two particular types of
connectivity distributions with easily tunable heterogeneity, namely,
networks with $\langle k\rangle N/2$ edges randomly distributed with $%
p\left( k\right) $ such that the heterogeneity depends on a single
parameter. Our first case is the bimodal distribution, $p(k)=\frac{1}{2}%
\delta (k-k_{1})+\frac{1}{2}\delta (k-k_{2})$ with parameter $\Delta
=(k_{2}-k_{1})/2=\langle k\rangle -k_{1}=k_{2}-\langle k\rangle $. Our
second case is the \textit{scale--free} distribution, $p(k)\sim k^{-\gamma
}, $ which does not have any characteristic size but $k$ is confined to the
limits, $k_{0}$ and $k_{m}\leq \min (k_{0}N^{\frac{1}{\gamma -1}},N-1)$ for
finite $N$ \cite{footnote2}. Notice that the network in this case gets more
homogeneous as $\gamma $ is increased \cite{footnote3}, and that this kind
of distribution seems to be most relevant in nature \cite%
{Barabasi,Torres_influence,nets2,nets3}. In particular, the \textit{%
functional topology} in the human brain, as defined by correlated activity
between small clusters of neurons, has been shown to correspond to this case
with exponent $\gamma \simeq 2$ \cite{Eguiluz} (in spite of the fact that
the brain's \textit{structural} or wiring topology is not yet well known).

\begin{figure}[h!]
\begin{center}
\psfig{file=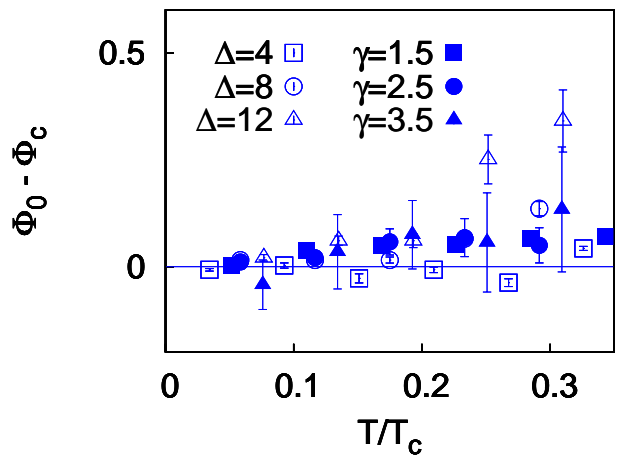, width=9cm}
\end{center}
\caption{The temperature
dependence of the difference between the values for the fatigue at which the
ferromagnetic--periodic transition occurs, as obtained analytically for $T=0$
($\Phi _{0}$) and from MC simulations at finite $T$ ($\Phi _{c}$). The
critical temperature is calculated as $T_{c}=\langle k^{2}\rangle \left(
\left\langle k\right\rangle N\right) ^{-1}$ for each topology. Data are for
bimodal distributions with varying $\Delta $ and for scale--free topologies
with varying $\protect\gamma ,$ as indicated. Here, $\langle k\rangle =20$, $%
N=1600$ and $\protect\alpha =2.$ The bars are standard deviations for 10
network realisations.}
\label{fig_Tc}
\end{figure}

We obtained the critical value of the fatigue, $\Phi _{c}\left( T\right) ,$
from Monte Carlo simulations at finite temperature $T.$ These indicate that
chaos never occurs for $T\gtrsim 0.35T_{c}.$ On the other hand, a detailed
comparison of the value $\Phi _{c}$ with $\Phi _{0}$ ---as obtained
analytically for $T=0$--- indicates that $\Phi _{c}\simeq \Phi _{0}.$ Figure %
\ref{fig_Tc}
illustrates the \textquotedblleft error\textquotedblright\ $\Phi
_{0}-\Phi _{c}\left( T\right) $ for different topologies. This shows that
the approximation $\Phi _{c}\simeq \Phi _{0}$ is quite good at low $T$ for
any of the cases examined. Therefore, assuming the critical values for the
main parameters, $T_{c}$ and $\Phi _{0},$ as given by our map, we conclude
that the more heterogeneous the distribution of connectivities of a network
is, the lower the amount of fatigue, and the higher the critical
temperature, needed to destabilize the dynamics. As an example of this
interesting behavior, consider a network with $\langle k\rangle =\ln (N)$,
and dynamics according to $\alpha =2$. If the distribution were regular, the
critical values would be $T_{c}=\ln (N)/N$ (which goes to zero in the
thermodynamic limit) and $\Phi _{0}=0$. However, a scale--free topology with
the same number of edges and $\gamma =2$ would yield $T_{c}=1$ and $\Phi
_{0}=1-2(\ln N)^{3}/N^{2}$ (which goes to 1 as $N\rightarrow \infty ).$

\begin{figure}[h!]
\begin{center}
\psfig{file=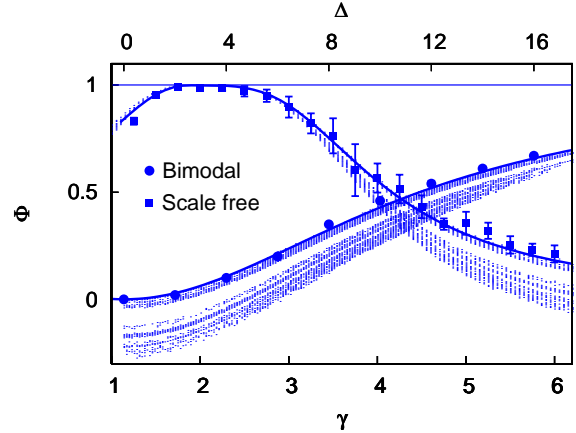, width=8.1cm}
\end{center}
\caption{The
critical fatigue values $\Phi _{0}$ (solid lines) and $\Phi _{c}$ from MC
averages over 10 networks (symbols) with $T=2/N,$ $\langle k\rangle =20,$ $%
N=1600,$ $\protect\alpha =2$. The dots below the lines correspond to changes
of sign of the Lyapunov exponent as given by the iterated map, which roughly
agree with the other results. This is for bimodal and scale--free
topologies, as indicated.}
\label{fig_phase}
\end{figure}

Figure \ref{fig_phase}
illustrates, for two topologies, the phase diagram of the
ferromagnetic--chaotic transition. Most remarkable is the plateau observed
in the \textit{edge-of-chaos} or transition\ curve for scale--free
topologies around $\gamma \simeq 2$, for which very little fatigue, namely, $%
\Phi \lesssim 1$ which corresponds to slight \textit{depression}, is
required to achieve chaos. The limit $\gamma \rightarrow \infty $
corresponds to $\langle k\rangle $--regular graphs (equivalent to $\Delta =0$%
). If $\gamma $ is reduced, $k_{m}$ increases and $k_{0}$ decreases. The
network is truncated when $k_{m}=N$. It follows that a value of $\gamma $
exits at which $k_{0}$ cannot be smaller, so that $k_{m}$ must drop to
preserve $\langle k\rangle $. This explains the fall in $\Phi _{c}$ as $%
\gamma \rightarrow 1$.

\begin{figure}[h!]
\begin{center}
\psfig{file=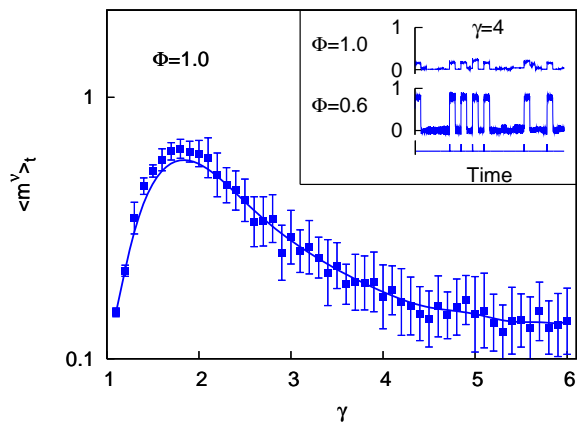, width=8.1cm}
\end{center}
\caption{Network \textquotedblleft performance\textquotedblright\ 
(see the main text) against 
$\protect\gamma $ for scale--free topology with $\Phi =1,$ as an average
over $20$ network realizations with stimulation every $50$ MC steps for $%
2000 $ MC steps, $\protect\delta =5$ and $M=4;$ other parameters as in Fig. 
\protect\ref{fig_phase}. Inset shows sections of typical time series of $m^{%
\protect\nu }$ for $\protect\gamma =4;$ the corresponding stimulus for
pattern $\protect\nu $ is shown below.}
\label{fig_stim}
\end{figure}

As a further illustration of our findings, we monitored the performance as a
function of topology during a simulation of pattern recognition. That is, we
\textquotedblleft showed\textquotedblright\ the system a pattern, say $\nu $
chosen at random from the set of $M$ previously stored, every certain number
of time steps. This was performed in practice by changing the field at each
node for one time step, namely, $h_{i}\rightarrow h_{i}+\delta \xi ^{\nu }$,
where $\delta $ measures the intensity of the input signal. Ideally, the
network should remain in this configuration until it is newly stimulated.
The performance may thus be estimated from a temporal average of the overlap
between the current state and the input pattern, $\langle m^{\nu }\rangle
_{time}$. This is observed to simply increase monotonically with $\Delta $
for the bimodal case. The scale--free case, however, as illustrated in
figure \ref{fig_stim}, 
shows how the task is better performed the closer to the edge of
chaos the network is. This is because the system is then easily destabilized
by the stimulus while being able to retrieve a pattern with accuracy. Figure %
\ref{fig_stim} also shows that the best performance for the scale--free
topology when $\Phi =1,$ i.e., in the absence of any fatigue, definetely
occurs around $\gamma =2.$%

The fact that the model network above is one of the simplest situations one
may conceive ---with dynamic connections which depend on local fields---
suggest that topological heterogeneity may indeed be a relevant property for
a complex network to perform efficiently certain high level functions. This
has in practice been illustrated before using networks with a similar
philosophy which happen to be useful for pattern recognition and class
identification \cite{class}. Our system retrieves memory patterns with
accuracy in spite of noise, and yet it may easily destabilize itself to
change state in response to an input signal ---without requiring an
excessive \textit{fatigue} for the purpose. There is a correlation between
the amount $\Phi $ of fatigue and the value of $\gamma $ for which
performance is maximized. One may argue that the plateau of
\textquotedblleft good\textquotedblright\ behavior shown around $\gamma
\simeq 2$ for scale--free networks with $\Phi \lesssim 1$ (figure \ref%
{fig_phase}) is a possible justification for the supposed tendency of
certain systems in nature to evolve towards this topology. It may also serve
as a hint when implementing artificial networks.%
%

We thank M.A. Mu\~{n}oz for very helpful comments. This work was financed by
the JA project FQM--01505 and by the MEC--FEDER project FIS2005--00791.


\begin{thebibliography}{99}
\bibitem{footnote1} Assuming one edge at most between any two nodes, $%
\epsilon _{ij}=0,1,$ the value will be slightly smaller, but it is easy to
prove that this is also a good approximation for $k_{i},k_{j}\ll N$. If more
than one edge is permitted, which is more realistic, this result is exact
for any random network, and it agrees (for a power--law distribution with $%
\gamma =3)$ with the one obtained in \cite{Bianconi} for an evolving network 
\cite{Barabasi}.
\bibitem{footnote2} Though the minimum connectivity is $1$, we shall assume $%
k_{0}\geq 2$, which reduces the probability of disconnection.
\bibitem{footnote3} The distribution is truncated and therefore not strictly
scale free for $\gamma <2$. However, nature shows examples for which $\gamma 
$ is slightly larger than 1, so that we consider the whole range here.
\end{thebibliography}
\end{document}